\documentclass[a4paper,12pt]{article}

\usepackage{amsmath}
\usepackage[psamsfonts]{amssymb}
\usepackage{rsfs}
\usepackage{bm}

\usepackage{cite}
\usepackage[dvips]{graphicx}
\begin{document} 

\begin{titlepage}

\baselineskip 10pt
\hrule 
\vskip 5pt
\leftline{}
\leftline{Chiba Univ. Preprint
          \hfill   \small \hbox{\bf CHIBA-EP-136}}
\leftline{\hfill   \small \hbox{hep-th/0209236}}
\leftline{\hfill   \small \hbox{September 2002}}
\vskip 5pt
\baselineskip 14pt
%\leftline{}
\hrule 
\vskip 1.0cm
\centerline{\Large\bf 
Infrared and ultraviolet asymptotic solutions 
} 
\vskip 0.5cm
\centerline{\Large\bf  
to gluon and ghost propagators 
}%$^*$
\vskip 0.5cm
\centerline{\Large\bf  
in Yang-Mills theory
}%$^*$
\vskip 0.5cm
\centerline{\large\bf  
}%$^*$

\vskip 0.5cm

\centerline{{\bf 
Kei-Ichi Kondo$^{\dagger,{1}}$ 
}}  
\vskip 0.5cm
\centerline{\it
${}^{\dagger}$Department of Physics, Faculty of Science, 
Chiba University, Chiba 263-8522, Japan
%$\&$ Graduate School of Science and Technology, 
}
%\vskip 0.3cm
%\centerline{\it
%Chiba University, Chiba 263-8522, Japan
%}
\vskip 1cm
%\begin{description}
%\item[]{\it \centerline{  
%${}^{\dagger}$Department of Physics, Faculty of Science, 
%Chiba University,  Chiba 263-8522, Japan}
%}
%\vskip -0.5cm
%\item[]{\it 
%${}^{\ddagger}$Graduate School of Science and Technology, 
%Chiba University, Chiba 263-8522, Japan
%}
%\end{description}

\begin{abstract}
We examine the possibility that there may exist a logarithmic correction to the infrared asymptotic solution with power behavior which has recently been found for the gluon and Faddeev-Popov ghost propagators in the Landau gauge.  
We propose a new Ansatz to find a pair of solutions for the gluon and ghost form factors by solving the coupled Schwinger-Dyson equation under a simple truncation.
This Ansatz enables us to derive the infrared and ultraviolet asymptotic solutions simultaneously and to understand why the power solution and the logarithmic solution is possible only in the infrared and ultraviolet limit respectively. 
Even in the presence of the logarithmic correction, the gluon propagator vanishes and the ghost propagator is enhanced  in the infrared limit, and the gluon-ghost-antighost coupling constant has an infrared fixed point (but with a different $\beta$ function).  This situation is consistent with Gribov-Zwanziger confinement scenario and color confinement criterion of Kugo and Ojima.

\end{abstract}

\vskip 0.5cm
Key words: Infrared exponent, Schwinger-Dyson equation, Yang-Mills theory, fixed point,  

PACS: 12.38.Aw, 12.38.Lg 
\vskip 0.2cm
\hrule  
\vskip 0.2cm
${}^1$ 
  E-mail:  {\tt kondo@cuphd.nd.chiba-u.ac.jp}

\vskip 0.2cm  

\par 
\par\noindent
\vskip 0.5cm

%%%%%%%%%%%%%%%%%%%%%%%%%%%%%%
%\vskip 2cm  
%\hrule  
%\bigskip  
%\centerline
%{\bf CHIBA UNIVERSITY}  
%\vfill 

%\hrule  
\vskip 0.5cm
%\begin{description}
%\item[]{
%$^\ddagger$
%Address from March 1996 to December 1996.
%  On leave of absence from: \\
%  Department of Physics, Faculty of Science,
%  Chiba University, Chiba 263, Japan.
%  }
%\item[]{
$^*$ To be published in Phys. Lett. B.
% Submitted to .
% }  
%\end{description}

\newpage
%\hrule  
%%%%% Table of Contents %%%%%
\pagenumbering{roman}
%\tableofcontents
%%%%% Table of Contents %%%%%
%%%%%
%\baselineskip 23pt

\vskip 0.5cm  
%\hrule  

%%%%%%%%%%%%%%%%%%%%%%%%%%

%%%%%%%%%%%%%%%%%%%%%%%%%%

\end{titlepage}

%\newpage

\pagenumbering{arabic}

\baselineskip 14pt
%%%%%%%%%%%%%%%%%%%%%%%%%%%%%%%%%%%%%%%%%%%%%%%%%%%%%%%%%%%%%%%%%%%%%%
\section{Introduction}

Recent investigations for the infrared (IR) dynamics of Yang-Mills theory based on various methods  converge to a consistent picture for the gluon and Faddeev-Popov ghost propagators: 
In the framework of Euclidean $SU(N_c)$ Yang-Mills theory with the manifestly Lorentz covariant gauge fixing, the transverse gluon propagator in the Landau gauge vanishes in the IR limit $p^2 \rightarrow 0$,
\begin{equation}
  \lim_{p^2 \rightarrow 0} D_T(p^2)  = 0 ,
\end{equation}
while the Faddeev-Popov ghost propagator is enhanced in the IR limit:  
\begin{equation}
  \lim_{p^2 \rightarrow 0}  G_{gh}(p^2)  = \infty ,
\end{equation}
where $D_T(p^2)$ and $G_{gh}(p^2)$ are respectively defined by the gluon propagator in the Landau gauge and the ghost propagator with unbroken color symmetry as
\begin{subequations}
\begin{align}
  D_{\mu\nu}^{AB}(p) :=& \delta^{AB}D_T(p^2) P_{\mu\nu}^T(p) ,
\\
 G^{AB}(p) :=& \delta^{AB} G_{gh}(p^2)  ,
\end{align}
\end{subequations}
with $A,B=1, \cdots, N_c^2-1$ and the transverse projection operator,
$
  P_{\mu\nu}^T(p) := \delta_{\mu\nu} -{p_\mu p_\nu \over p^2} .
$

Such behaviors of the gluon and ghost propagators were first derived by Gribov \cite{Gribov78} long ago as a result of restricting the region of functional integration over the gluon field to the interior of the Gribov horizon in order to avoid Gribov copies. 
Indeed, he obtained the result:
\begin{equation}
  D_T(p^2) = {p^2 \over (p^2)^2+M^4}, \quad G_{gh}(p^2) \sim {M^2 \over (p^2)^2} ,
\end{equation}
where $M$ is a mass scale called the Gribov mass.  
Subsequently, Zwanziger \cite{Zwanziger91,Zwanziger94} has obtained the exact analytic result that the restriction of the Gribov horizon enforced by a horizon condition yields 
\begin{equation}
  \lim_{p^2 \rightarrow 0} D_T(p^2)  = 0 ,
  \quad
  \lim_{p^2 \rightarrow 0}  p^2 G_{gh}(p^2)  = \infty .
\end{equation}

Recent studies \cite{SHA97,AB98,AB98b,Bloch01,FAR02,LS02} of the coupled SD equation for the gluon and ghost propagators have reached a conclusion that, in the IR limit $p^2 \rightarrow 0$, the gluon form factor $F(p^2)$ and ghost form factor $G(p^2)$ defined by
\begin{align}
  D_T(p^2) :=    {F(p^2)/p^2}   ,
\quad
 G_{gh}(p^2) :=   {G(p^2)/p^2}  ,
\end{align}
exhibit the power behavior characterized by an IR critical exponent $\kappa$:
\begin{align}
  F(p^2) = A (p^2)^{2\kappa}, \quad G(p^2) = B (p^2)^{-\kappa} ,
\label{IRsol}
\end{align}
where $A$ and $B$ are $p$-independent constants, and $\kappa$ takes the value between 1/2 and 1 depending on the approximations adopted to write down the solvable SD equation (Gribov's result corresponds to $\kappa=1$.  However, the aim of this paper is not to obtain the precise value of $\kappa$.).
These IR asymptotic solutions should be compared with the ultraviolet (UV) asymptotic solutions obtained by perturbation theory:
\begin{align}
  F(p^2) = C (\log p^2)^\gamma , \quad G(p^2) = D (\log p^2)^\delta ,
\label{UVsol}
\end{align}
where $C$ and $D$ are $p$-independent constants and, at one-loop level, the exponents $\gamma$ and $\delta$ are independent of the number of colors $N_c$ in $SU(N_c)$ Yang-Mills theory, i.e., 
$\gamma=-{13 \over 22}$ and $\delta=-{9 \over 44}$.
This type of solutions with logarithmic dependence on momenta can  also be derived as the UV asymptotic solution to the SD equation, see \cite{SHA97,FAR02}.  

It is possible to explain how Gribov's old result is compatible with the  recent IR  solution of SD equation.  Gribov's prescription to cut off the functional integral at the first  Gribov horizon does not alter the SD equations of Faddeev-Popov theory, as pointed out by \cite{Zwanziger01}.  The cut-off at the first Gribov horizon assures that both the gluon and ghost propagators are positive, and that negative and   oscillating solutions are excluded.  Indeed, the above IR solutions for the truncated SD equations are positive $A, B>0$,

Moreover, the method of the stochastic quantization in its time-independent formulation indicates qualitatively the same result \cite{Zwanziger01,Zwanziger02} as the SD equation.
Recent numerical simulations on the lattice also support the above results, see e.g., 
\cite{Bonnetetal00,LRG02,NF02,BCLM02}. 

The above result has an implication to the confinement problem.
As an immediate consequence of the suppression of the gluon propagator in the IR limit, gluon confinement is easily understood, since the would-be physical transverse gluon becomes short range and the massless gluon is absent from physical spectrum.  On the other hand, infrared enhancement of ghost propagator suggests that the unphysical longitudinal gluon becomes long range and mediates the confining strong force, providing the linear confining potential, i.e., quark confinement. 
More generally, it is worth remarking that the enhancement of the ghost propagator is consistent with a sufficient condition for color confinement (confinement of all the color non-singlet objects) due to Kugo and Ojima \cite{KO79} (see \cite{WA01}):
\begin{equation}
  \lim_{p^2 \rightarrow 0} G^{-1}(p^2) \equiv 1+u(0) = 0 .
\end{equation}
Thus the precise information of the IR limit is quite important to elucidate the non-perturbative dynamics of Yang-Mills theory.

In this paper, we re-examine the truncated Schwinger-Dyson equation for the gluon and ghost propagators in Yang-Mills theory in the Landau gauge, keeping these developments in mind.
A purpose of paper is to point out that the above solution (\ref{IRsol}) is not the most general IR asymptotic solution of the truncated SD equation. 
Another purpose is to understand why the quite different asymptotic solutions, i.e., IR power solution and UV logarithmic solution in momentum space, can be obtained from the same SD equation.  
For this end, we study the IR and UV asymptotic solution on equal footing simultaneously by proposing a new Ansatz and examine a possible  logarithmic correction to the IR asymptotic power solution.

In the studies of SD equation so far \cite{SHA97,AB98,AB98b,Bloch01,FAR02,LS02}, an Ansatz for the solution, (\ref{IRsol}) or (\ref{UVsol}), was substituted into the SD equation and the power index and the coefficient are determined self-consistently by comparing both sides of SD equation.  
However, this procedure does not guarantee the uniqueness of the solution. 
Moreover, improving the IR asymptotic solution was attempted by von Smekal, Hauck and Alkofer \cite{SHA97} and Atkinson and Bloch \cite{AB98}.  
It has been shown that the truncated SD equation with bare vertex functions is satisfied by the asymptotic expansion (according to the notation \cite{AB98}): 
\begin{align}
 F(p^2) = A_0 (p^2)^{2\kappa} \left( 1 + \sum_{n=1}^{N} f_n a^n (p^2)^{\rho n} \right) ,
\quad
 G(p^2) = B_0 (p^2)^{-\kappa} \left( 1 + \sum_{n=1}^{N} g_n a^n (p^2)^{\rho n} \right),
\label{ABsolution}
\end{align}
where $\kappa=0.769479$ ($\rho=1.96964$ and $f_1=1,g_1=0.829602$) and $a$ is a free parameter which can not be determined from the IR asymptotic expansion alone. The IR parameter $a$ is expected to be negative for the IR asymptotic solution to connect into the UV asymptotic solution (in a very narrow transition region).  The value was determined by numerical solution to be negative, 
$a=-10.27685$, as expected.  See \cite{FAR02} for further information. 

In this paper, we adopt another Ansatz for asymptotic solutions which incorporates possible logarithmic dependence and can be applied to both IR region $p^2 \rightarrow 0$ and UV region $p^2 \rightarrow \infty$ simultaneously:
\footnote{In the IR region, we can add the power-series correction part as in (\ref{ABsolution}).  
Moreover, we can also treat the region $z \cong z_\sigma \ll 1$ ($p^2 \cong \sigma$) by changing the last series as
\begin{align}
 F(p^2) = A e^{\alpha z/\omega} z^\gamma \left( 1 + \sum_{n=1}^{N} c_n z^n \right) ,
\quad
 G(p^2) = B e^{\beta z/\omega} z^\delta \left( 1 + \sum_{n=1}^{N} d_n z^n \right), 
\label{Kansatz}
\end{align}
}
\begin{align}
 F(p^2) = A e^{\alpha z/\omega} z^\gamma \left( 1 + \sum_{n=1}^{N}{c_n \over z^n} \right) ,
\quad
 G(p^2) = B e^{\beta z/\omega} z^\delta \left( 1 + \sum_{n=1}^{N}{d_n \over z^n} \right),
\label{ansatz}
\end{align}
with a new variable \cite{KSY91},
\begin{align}
  z :=  \omega \ln {p^2 \over \sigma} + z_\sigma ,
\end{align}
where $\omega$ is an appropriate real number  and 
$\sigma$ is the renormalization group invariant momentum scale defined by
$\sigma:=\mu^2 \exp [-2 \int_{g_0}^{g} {d\lambda \over \beta(\lambda)}]$ for the renormalization scale $\mu$ using the $\beta$ function: 
$\beta(g):= \mu {dg(\mu) \over d\mu}$.
Here the coefficients $c_n$ and $d_n$ can be expressed in terms of $\gamma$ and $\delta$.  
The IR and UV asymptotic region corresponds to $|z| \rightarrow \infty$ where the signature of $z$ depends on that of $\omega$.
Note that these parameters are not all independent (We can give a certain value for $z_\sigma$, e.g., $z_\sigma=1$ or $z_\sigma=0$, since it is negligible when 
$|\ln {p^2 \over \sigma}| \rightarrow \infty$).  

An advantage of this Ansatz is that we can search for the possible IR and UV solutions simultaneously without discriminating the IR and UV region by specifying  possible set of power indices and coefficients.  
In fact, we show that the above Ansatz satisfies 
the truncated SD equation \cite{AB98} specified below, if the exponents satisfy either of the two relationships, 
\begin{subequations}
\begin{align}
  \alpha = 0 = \beta, \quad \gamma + 2\delta = -1, \cdots, 
\label{UV}
\\
  \alpha = - 2\beta \not= 0, \quad \gamma =- 2\delta  ,  \cdots, 
\label{IR}
\end{align}
\end{subequations}
where the relationships among $c_n, d_n$ are given later, since they are very complicated (see Appendix).  
It should be remarked again that these relations can be obtained without specifying the momentum region in question, by comparing the both sides of the respective SD equation.  We must choose the appropriate solution as follows.

The asymptotic UV solution is selected by requiring 
\begin{align}
  F(p^2),  G(p^2) \rightarrow 0 \quad \text{as} \quad p^2 \rightarrow \infty .
\end{align}
Therefore, the first choice (\ref{UV}) of parameters corresponds to the UV asymptotic solution: pure logarithmic solution without leading power behavior.  Indeed, the perturbative result $\gamma=-{13 \over 22}$ and $\delta=-{9 \over 44}$ satisfies $\gamma + 2\delta = -1$.  

Another advantage is that the IR asymptotic solution allows the logarithmic correction to the power solution.  
It is shown below that the IR solution is identified with the second set of parameters (\ref{IR}) ($\kappa=-\beta\not=0$):
\begin{align}
 F(p^2) = A e^{2\kappa z/\omega} z^{-2\delta}  \sum_{n=0}^{N} c_n z^{-n} ,
\quad
 G(p^2) = B e^{-\kappa z/\omega} z^\delta  \sum_{n=0}^{N} d_n z^{-n} ,
\label{Kondosolution}
\end{align}
where $c_0=1=d_0$ and the coefficient $c_n, d_n$ can now be expressed by $\delta$.
Irrespective of the value of $\delta$, these solutions  
for $\kappa>0$ behave  as
\begin{align}
  F(p^2) \sim (p^2)^{2\kappa} (-\log p^2)^{-2\delta} \rightarrow 0 , \quad 
G(p^2) \sim (p^2)^{-\kappa} (-\log p^2)^{\delta} \rightarrow \infty   .
\end{align}
For $\kappa>1/2$, in particular, we have the IR suppressed vanishing gluon propagator and IR enhanced ghost propagator, irrespective of the presence or absence of the logarithmic correction.  Therefore, the existence of logarithmic correction does not change the expected IR physics at the IR limit $p^2 \rightarrow 0$.
Thus the new type of solution, even if it exists, does not destroy the argument in favor of color confinement mentioned above.
If $\delta=0$ and $c_n=0=d_n$ for all $n\ge 1$, the solution in question reduces to the pure power solution known so far.

The identification of the IR solution is done as follows.
First of all, it should be remarked that all the solutions of SD equations are not allowed as physically meaningful solutions. 
To clarify this point, we examine the renormalized (running) gluon-ghost-antighost coupling constant defined by
\begin{align}
  g^2(p) := g^2 F(p^2)G^2(p^2) ,
\end{align}
where $g^2(p)$ is renormalization group invariant in the Landau gauge, since $Z_g Z_3^{1/2} \tilde{Z}_3 = \tilde{Z}_1 =1$. 
The above Ansatz yields the running coupling
$
 g^2(p) = g^2 AB^2 e^{(\alpha+2\beta) z/\omega} z^{\gamma+2\delta} 
$
$
\times \left( 1 + \sum_{n=1}^{N} c_n z^{-n} \right) 
 \left( 1 + \sum_{n=1}^{N} d_n z^{-n} \right)^2 .
$
The first set of parameters leads to
\begin{align}
 g^2(p) = g^2 AB^2 z^{-1}  \left( 1 +  (c_1+2d_1)z^{-1} + O\left(z^{-2}\right) \right) ,
\label{rc1}
\end{align}
which vanishes in both UV $p^2 \rightarrow \infty$ and IR limit $p^2 \rightarrow 0$.  
To be consistent with the UV asymptotic freedom, the running coupling must become zero at $p=\infty$ and grows as $p^2$ decreases.  Therefore, the running coupling (\ref{rc1}) can be identified with the UV solution.  However, if we adopt this solution also as the IR solution, the running coupling must vanish at $p^2=0$ again.  This identification leads to the multi-valued $\beta$ function.  Hence, the first set can not correspond to the IR solution.

\begin{figure}[htbp]
\begin{center}
\includegraphics{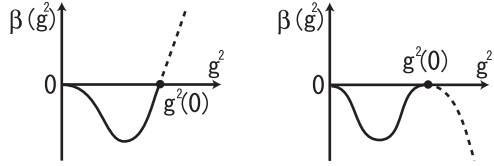}
\caption{The $\beta$ functions $\beta(g^2)$ with an infrared fixed point  at $g^2(0)$ corresponding to the IR power solution in the absence (left) and presence (right) of logarithmic correction.}
\label{fig:betafunc}
\end{center}
\end{figure}

On the other hand, the second choice  
\begin{align}
 g^2(p) = g^2 AB^2 \left( 1 +  (c_1+2d_1)z^{-1} + O\left( z^{-2}\right) \right)  
\end{align}
converges to a finite and non-zero value in both UV $p^2 \rightarrow \infty$ and IR limit $p^2 \rightarrow 0$ irrespective of $\beta, \delta$.  
Hence this is not accepted as the UV solution.  
Identification with the IR solution implies that $g^2(0)=g^2AB^2$ is nothing but the IR stable fixed point for arbitrary values of $\kappa, \delta$. 
As $p$ increases towards the UV region from the IR fixed value at $p=0$, therefore, the running coupling must bend below to smoothly connect into the UV logarithmic solution.  This effect can be seen by taking into account the power series in $z^{-1}$.  For this purpose, the parameters $c_1$ and $d_1$ must satisfy $(c_1+2d_1)/z<0$, provided that $c_1 d_1 \not= 0$. 
Incidentally, the IR diverging running coupling constant does not appear from the solutions subject to the above relationship of parameters.  

The UV behavior is characterized by the $\beta$ function, 
$\beta(g^2)=-{11N_c/3 \over 8\pi^2} g^4 +o(g^6)$
with $\beta'(0)=0$.  
In the neighborhood of IR fixed point $g^2(0)$, 
the beta function 
$\beta(g^2) := \mu {dg^2(\mu) \over d\mu}$
behaves as follows (See Fig.~\ref{fig:betafunc}). 
The power solution with power correction (\ref{ABsolution}) leads to
$
 \beta(g^2)  
= 2\rho g(0)^2 (g^2/g^2(0)-1) + o((g^2/g^2(0)-1)^2) ,
$  
and 
$\beta'(g^2(0))=2\rho>0$,
while the power solution with logarithmic correction (\ref{Kondosolution}) yields
$
 \beta(g^2) = -{2\omega g(0)^2 \over c_1+d_1}(g^2/g^2(0)-1)^2 
+ o((g^2/g^2(0)-1)^3) , 
$
and 
$\beta'(g^2(0))=0$. 
The difference can not be seen in the $g^2(p)$ vs. $\ln p$ plot, since ${d g^2(p) \over d\ln p}=0$ at $p=0$ in both cases.
However, in the $g^2(p)$ vs. $p$ plot, the logarithmic correction implies
${d g^2(p) \over dp}=-\infty$ at $p=0$,  
in contrast to the power solution 
${d g^2(p) \over dp}=0$ at $p=0$.
The data of numerical simulation available seem to be still insufficient to discriminate between two cases.  
In this paper, therefore, we examine the IR behavior in the analytical way.

\section{Finding asymptotic solutions of the coupled SD equation}

\begin{figure}[htbp]
\begin{center}
\includegraphics{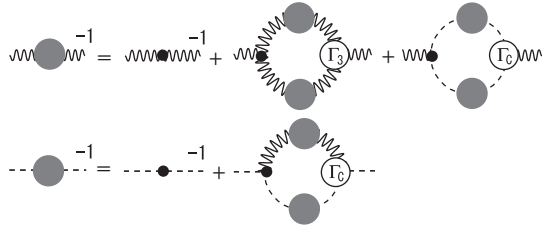}
\caption{Diagrammatic representation of the truncated Schwinger-Dyson equations for gluon and ghost propagators.}
\label{fig:SDeq}
\end{center}
\end{figure}

For simplicity of analysis, we restrict our consideration to a version of the truncated SD equation adopted by 
Atkinson and Bloch \cite{AB98} using the bare vertex functions, since
qualitatively the same IR behavior was obtained also from the solutions of the truncated SD equations using the full vertex function improved so as to be consistent with the Slavnov-Taylor identity.

The truncation is done as follows. 
We include all diagrams in the ghost equation, while we neglect contributions from the two-loop diagrams in the gluon equation.  
The gluon equation was contracted with the Brown-Pennington projection operator \cite{BP88} 
$
  R_{\mu\nu}(p) := \delta_{\mu\nu} - 4{p_\mu p_\nu \over p^2} ,
$
which removes the tadpole term.  Therefore, the quadratic UV divergence can be removed by this projection.
Finally, the non-trivial contribution in the gluon equation comes from only two diagrams, i.e., gluon loop and ghost loop, apart from the tree gluon propagator.  We include only the bare triple gluon vertex and the bare gluon-ghost-antighost vertex. 
See Fig.~\ref{fig:SDeq}.

Moreover, in order to avoid angular integration, we adopt the approximation called the $y$-max approximation, 
$F((p-q)^2)=F(\text{max}\{p^2,q^2\})$,
$G((p-q)^2)=G(\text{max}\{p^2,q^2\})$.
Here the angle $\theta$ comes from the inner product $p \cdot q= \sqrt{p^2} \sqrt{q^2} \cos \theta$ between the external momenta $p$ and the internal momenta $q$ which is to be integrated out.
Thus we can write explicitly down the coupled SD equation each of which is the one-dimensional integral equation.%
\footnote{This approximation was also called Landau-Abrikosov-Khalatnikov (LAK) approximation or Higashijima-Miransky (HM) \cite{HM84} approximation.  
In the bare vertex approximation,  the exact angular integration was performed for the pure power solution \cite{AB98b} without relying on this approximation, 
which yields a largest value of $\kappa$, i.e., $\kappa=1$.
This value agrees with the Gribov prediction.
Recently, it has been shown \cite{FAR02,LS02} that the value of $\kappa$ is decreased between 0.5 and 0.6 due to the renormalization effect of the triple gluon interaction vertex which is necessary to reproduce the correct coefficient of $\beta$ function in the UV limit. 
}

Introducing the renormalization constants for the gluon field $Z_3$, the ghost field $\tilde{Z}_3$, the triple gluon vertex $Z_1$ and the gluon-ghost-antighost vertex $\tilde{Z}_1$, 
we write down the SD equation for the renormalized gluon and ghost form factors.  
By introducing 
$\lambda :={g^2 \over 16\pi^2}$, $x:=p^2$ and $y:=q^2$,
the ghost SD equation reads
\begin{align}
  G^{-1}(x) = \tilde{Z}_3 - {3N_c \over 4}\lambda \tilde{Z}_1 \Big[ {F(x) \over x^2} \int_{0}^{x} dy y G(y) + \int_{x}^{\Lambda^2} {dy \over y} F(y) G(y) \Big] ,
\end{align}
while the gluon SD equation reads
\begin{align}
  F^{-1}(x) =& {Z}_3 + {N_c \over 3}\lambda \tilde{Z}_1 \Big[ -{G(x) \over x^3} \int_{0}^{x} dy y^2 G(y) + {3G(x) \over 2x^2} \int_{0}^{x} dy y G(y) 
  + {1 \over 2} \int_{x}^{\Lambda^2} {dy \over y} G^2(y)  \Big] 
\nonumber\\
&+ {N_c \over 3}\lambda {Z}_1 \Big[  {7F(x) \over 2x^3} \int_{0}^{x} dy y^2 F(y) 
- {17F(x) \over 2x^2} \int_{0}^{x} dy y F(y) 
- {9F(x) \over 8x} \int_{0}^{x} dy F(y) 
\nonumber\\ \quad \quad 
  &- 7 \int_{x}^{\Lambda^2} {dy \over y} F^2(y)
  + {7x \over 8} \int_{x}^{\Lambda^2} {dy \over y^2} F^2(y)  \Big] .
\end{align}
The renormalization constants $Z_3$ and $\tilde{Z}_3$ are eliminated by subtracting the equation at $x=\sigma$.
The ghost equation reads
\begin{align}
  G^{-1}(x) = G^{-1}(\sigma) - {3N_c \over 4}\lambda \tilde{Z}_1 \Big[ {F(x) \over x^2} \int_{0}^{x} dy y G(y) 
  - (x \rightarrow \sigma)
   + \int_{x}^{\sigma} {dy \over y} F(y) G(y) \Big] ,
\end{align}
while the gluon equation reads
\begin{align}
  F^{-1}(x) =& F^{-1}(\sigma) + {N_c \over 3}\lambda \tilde{Z}_1 \Big[ -{G(x) \over x^3} \int_{0}^{x} dy y^2 G(y) + {3G(x) \over 2x^2} \int_{0}^{x} dy y G(y) 
\nonumber\\
  & -(x \rightarrow \sigma)
  + \int_{x}^{\sigma} {dy \over 2y} G^2(y)  \Big] 
\nonumber\\
&+ {N_c \over 3}\lambda {Z}_1 \Big[  {7F(x) \over 2x^3} \int_{0}^{x} dy y^2 F(y) 
- {17F(x) \over 2x^2} \int_{0}^{x} dy y F(y) 
- {9F(x) \over 8x} \int_{0}^{x} dy F(y) 
\nonumber\\
  &+ {7x \over 8} \int_{x}^{\Lambda^2} {dy \over y^2} F^2(y) 
  - (x \rightarrow \sigma)
  - 7 \int_{x}^{\sigma} {dy \over y} F^2(y) \Big]  .
\end{align}

First, we rewrite the coupled SD equation in terms of the new variable $z:=z_\sigma+\omega \ln {x \over \sigma}$ and 
$\zeta:=z_\sigma+\omega \ln {y \over \sigma}$.  
The ghost equation reads
\begin{align}
  G^{-1}(z) - G^{-1}(z_\sigma) =& 
  - {3N_c \over 4}\lambda \tilde{Z}_1 \Big[  F(z) e^{-2z/\omega} \int_{-\omega \infty}^{z} {d\zeta \over \omega} e^{2\zeta/\omega} G(\zeta) 
\nonumber\\&
   - (x \rightarrow \sigma; z \rightarrow z_\sigma)
  + \int_{z}^{z_\sigma} {d\zeta \over \omega} F(\zeta) G(\zeta) \Big] ,
\end{align}
and the gluon equation reads
\begin{align}
  & F^{-1}(z) - F^{-1}(z_\sigma) 
\nonumber\\ 
=& 
    {N_c \over 3}\lambda \tilde{Z}_1 \Big[ - G(z) e^{-3z/\omega} \int_{-\omega \infty}^{z} {d\zeta \over \omega} e^{3\zeta/\omega} G(\zeta) 
  + {3 \over 2} G(z) e^{-2z/\omega} \int_{-\omega \infty}^{z} {d\zeta \over \omega} e^{2\zeta/\omega} G(\zeta) 
\nonumber\\
  & -(x \rightarrow \sigma; z \rightarrow z_\sigma)
  + {1 \over 2} \int_{z}^{z_\sigma} {d\zeta \over \omega}  G^2(\zeta)  \Big] 
\nonumber\\
&+ {N_c \over 3}\lambda {Z}_1 \Big[  {7 \over 2} e^{-3z/\omega} F(z) \int_{-\omega \infty}^{z} {d\zeta \over \omega} e^{3\zeta/\omega} F(\zeta) 
- {17 \over 2} e^{-2z/\omega} F(z) \int_{-\omega \infty}^{z} {d\zeta \over \omega} e^{2\zeta/\omega} F(\zeta) 
\nonumber\\&
- {9 \over 8} e^{-z/\omega} F(z) \int_{-\omega \infty}^{z} {d\zeta \over \omega} e^{\zeta/\omega} F(\zeta)  
  + {7 \over 8} e^{z/\omega} \int_{z}^{z_\Lambda} {d\zeta \over \omega} e^{-\zeta/\omega} F^2(\zeta) 
\nonumber\\&
  - (x \rightarrow \sigma; z \rightarrow z_\sigma)
  - 7 \int_{z}^{z_\sigma} {d\zeta \over \omega} F^2(\zeta) \Big] .
\end{align}

Next, we substitute the Ansatz (\ref{ansatz}) into the above equation \footnote{In order to use the asymptotic solution in the integrand, we perform the decomposition: 
$\int_{x}^{\Lambda^2}dy f(y) = - \int_{\epsilon^2}^{x}dy f(y) + \int_{\epsilon^2}^{\Lambda^2}dy f(y)$
 for the IR limit $x \rightarrow \epsilon (\rightarrow 0)$ and 
$\int_{0}^{x}dy f(y) =  \int_{\Lambda^2}^{x}dy f(y) + \int_{0}^{\Lambda^2}dy f(y)$
for the UV limit $x \rightarrow \Lambda^2 (\rightarrow \infty)$. 
In our approach, the term 
$\int_{0}^{\Lambda^2}dy f(y)$ can not be determined, but it is a $p$-independent constant and does not affect the exponent.  
}
and perform the integration over $\zeta$ by making use of the integration formula \cite{GR00}
\begin{align}
 \int^z d\zeta e^{a\zeta} \zeta^{b} = \begin{cases}
 (-1)^{-b} a^{-(1+b)} \Gamma[1+b, -az] & \text{($a \not=0$)} \\ 
 (1+b)^{-1} z^{1+b} & \text{($a=0$)} 
 \end{cases} ,
\end{align}
where $\Gamma[c,x]$ is the incomplete gamma function with the asymptotic expansion for large $|x|$:
$
%\begin{equation}
 \Gamma[c,x] = x^{c-1} e^{-x} \left[ 1 + \sum_{n=1}^{\infty} {(c-1)(c-2) \cdots (c-n) \over x^n} \right] .
%\end{equation}
$
In particular, for large $z$ and $a\not=0$, we can use the formula,
\begin{align}
 \int^z d\zeta e^{a\zeta} \zeta^{b} =  
  a^{-1}  e^{az} z^{b} \left[ 1 + \sum_{n=1}^{\infty} {b(b-1) \cdots (b+1-n) \over a^n} { (-1)^n \over z^n} \right] . 
\end{align}

The ghost equation is integrated out, when $\alpha\not=-\beta$ and $\beta\not=-2$:
\begin{align}
  & B^{-1} e^{-\beta z/\omega}(z^{-\delta}-d_1 z^{-\delta-1}+\cdots) 
  - ([z \rightarrow z_\sigma])
\nonumber\\
  =& 
  - {3N_c \over 4}\lambda \tilde{Z}_1 AB 
  \Big[  {e^{(\alpha+\beta)z/\omega} \over 2+\beta}(z^{\gamma}+c_1 z^{\gamma-1} +o(z^{\gamma-2})) 
      \left\{  z^{\delta}-{\omega \delta \over 2+\beta} z^{\delta-1} + d_1  z^{\delta-1}+o(z^{\delta-2}) \right\}
\nonumber\\&
  - {e^{(\alpha+\beta)z/\omega} \over \alpha+\beta} 
\left\{  z^{\gamma+\delta}-{\omega(\gamma+\delta) \over \alpha+\beta} z^{\gamma+\delta-1}  +(c_1+d_1) z^{\gamma+\delta-1} +o(z^{\gamma+\delta-2}) \right\} 
-  ([z \rightarrow z_\sigma])\Big] ,
\end{align}
where we have required $\beta>-2$ for eliminating the terms coming from the lower bound $-\omega \infty$ to avoid the IR singularity.   
By comparing both sides of this equation, we find that the following  relations must be satisfied. 
\begin{subequations}
\begin{align}
  \alpha + 2\beta &= 0, 
  \label{ghost1}
\\
  \gamma + 2\delta &= 0,
  \label{ghost2}
\\
  1&= {3N_c \over 4}\lambda \tilde{Z}_1 AB^2 \left( -{1 \over 2+\beta}+{1 \over \alpha+\beta} \right) ,
  \label{ghost3}
\\
 d_1 &= {3N_c \over 4}\lambda \tilde{Z}_1 AB^2 \left[ {c_1+d_1-{\omega \delta \over 2+\beta} \over 2+\beta} - {c_1+d_1-{\omega(\gamma+\delta) \over \alpha+\beta} \over \alpha+\beta} \right] .
  \label{ghost4}
\end{align}
\end{subequations}

Similarly, the gluon equation reads, 
if $\alpha\not=1/2, 0,-1, -2, -3$ and $\beta\not=0,-2,-3$:
\begin{align}
  & A^{-1}e^{-\alpha z/\omega}(z^{-\gamma}-c_1 z^{-\gamma-1} + o(\zeta^{2\gamma-2})) -([z \rightarrow z_\sigma]) 
\nonumber\\ 
=& 
    {N_c \over 3}\lambda \tilde{Z}_1 B^2 \Big[ -  {e^{2\beta z/\omega} \over 3+\beta} (z^{\delta}+d_1 z^{\delta-1}+o(\zeta^{2\gamma-2}) )  
    \left\{ z^{\delta}-{\omega \delta \over 3+\beta}z^{\delta-1} +d_1 z^{\delta-1} + o(\zeta^{2\gamma-2}) \right\} 
\nonumber\\&
  + {3 \over 2} {e^{2\beta z/\omega} \over 2+\beta} (z^{\delta}+d_1 z^{\delta-1}+o(\zeta^{2\gamma-2}) ) 
  \left\{ z^{\delta}-{\omega \delta \over 2+\beta}z^{\delta-1} +d_1 z^{\delta-1} + o(\zeta^{2\gamma-2}) \right\}
\nonumber\\
  & 
  - {1 \over 2} {e^{2\beta z/\omega} \over 2\beta} \left\{ z^{2\delta}-{\omega \delta \over \beta}z^{2\delta-1}+2d_1 z^{2\delta-1} + o(\zeta^{2\gamma-2}) \right\} 
-([z \rightarrow z_\sigma])  \Big] 
\nonumber\\
&+ {N_c \over 3}\lambda {Z}_1 A^2 \Big[  
{7 \over 2}  {e^{2\alpha z/\omega} \over 3+\alpha} (z^{\gamma}+c_1 z^{\gamma-1}+o(\zeta^{2\gamma-2}) )
  \left\{ z^{\gamma}-{\omega \gamma \over 3+\alpha} z^{\gamma-1} +c_1 z^{\gamma-1} + o(\zeta^{2\gamma-2}) \right\}   
\nonumber\\&
- {17 \over 2} {e^{2\alpha z/\omega} \over 2+\alpha} (z^{\gamma}+c_1 z^{\gamma-1}+o(\zeta^{2\gamma-2}) )
 \left\{ z^{\gamma}-{\omega \gamma \over 2+\alpha} z^{\gamma-1} +c_1 z^{\gamma-1} + o(\zeta^{2\gamma-2}) \right\} 
\nonumber\\&
- {9 \over 8} {e^{2\alpha z/\omega} \over 1+\alpha} (z^{\gamma}+c_1 z^{\gamma-1}+o(\zeta^{2\gamma-2}) )
 \left\{ z^{\gamma}-{\omega \gamma \over 1+\alpha} z^{\gamma-1} +c_1 z^{\gamma-1} + o(\zeta^{2\gamma-2}) \right\}
\nonumber\\&
  - {7 \over 8} {e^{2\alpha z/\omega} \over 2\alpha-1}  \left\{ z^{2\gamma}-{2\omega \gamma\over 2\alpha-1}z^{2\gamma-1}+2c_1 z^{2\gamma-1} + o(\zeta^{2\gamma-2}) \right\}  
+ o(e^{z/\omega})
\nonumber\\&
  + 7 {e^{2\alpha\zeta/\omega} \over 2\alpha} \left\{ z^{2\gamma}-{\omega \gamma \over \alpha}z^{2\gamma-1}+2c_1 z^{2\gamma-1} + o(\zeta^{2\gamma-2}) \right\} 
- ([z \rightarrow z_\sigma])  \Big]  ,
\label{gluoninteg2}
\end{align}
where we have required $\alpha>-2$ and $\beta>-2$ for eliminating the terms coming from the lower bound $-\omega \infty$.
In the right hand side of (\ref{gluoninteg2}), the former half is the contribution from ghost loop and the latter one comes from gluon loop.  
If $\alpha>0$, then $\beta=-\alpha/2<0$ and hence the ghost loop is dominant over the gluon loop.  This phenomenon was called the ghost dominance in the pure power solution.  
We confirm that the ghost dominance also holds even in presence of logarithmic correction (under the truncation adopted).   

The parameters must satisfy the following relations.
\begin{subequations}
\begin{align}
  \alpha + 2\beta &= 0 , 
  \label{gluon1}
\\ 
  \gamma + 2\delta &= 0,
  \label{gluon2}
\\ 
 1 &= {N_c \over 3}\lambda \tilde{Z}_1 AB^2 \left[ -{1 \over 3+\beta} +{3 \over 2}{1 \over 2+\beta} -{1 \over 4\beta} \right] ,
 \label{gluon3}
\\ 
 & c_1 + d_1 {N_c \over 3}\lambda \tilde{Z}_1 AB^2 \left( -{2 \over 3+\beta}+{3 \over 2+\beta}-{1 \over 2\beta} \right) 
 \nonumber\\ 
  &=  -{N_c \over 3}\lambda \tilde{Z}_1 AB^2 \left[ {1 \over (3+\beta)^2}-{3 \over 2}{1 \over (2+\beta)^2}+{1 \over 4\beta^2} \right] \omega \delta .
  \label{gluon4}
\end{align}
\end{subequations}
The relation (\ref{ghost1}) is equal to  (\ref{gluon1}), and
the relation (\ref{ghost2}) is equal to  (\ref{gluon2}).
 From the relation (\ref{ghost3}) and  (\ref{gluon3}), we obtain
\begin{align}
 \nu := \lambda \tilde{Z}_1 AB^2 
 = {1 \over N_c} 3 \left( -{2 \over 3+\beta}+{3 \over 2+\beta}-{1 \over 2\beta} \right)^{-1}
 = {1 \over N_c} {4 \over 3} \left( {1 \over 2+\beta}+{1 \over \beta} \right)^{-1} .
\end{align}
Therefore, $\beta$ satisfies the algebraic equation,
$
 19 \beta^2 + 77 \beta + 48 = 0 ,
$
whose solutions are 
$
 \beta = {77 \pm \sqrt{2281} \over 38} \cong - 0.769479, -3.28315 ,
$
irrespective of $N_c$. 
By adopting $\beta=-0.769479 (>-2)$, the IR fixed point of strong fine structure constant is given by
\begin{equation}
 \alpha_s(0) := 4\pi \lambda \tilde{Z}_1 AB^2 
 = {(g^2/4\pi)} \tilde{Z}_1 AB^2 
 = {(3/N_c)} 11.4702 .
\end{equation}
The running coupling in the neighborhood of $p^2=0$ is given by
\begin{equation}
 \alpha_s(p)  
 = {(g^2/4\pi)} \tilde{Z}_1 F(p^2) G^2(p) 
 = {(3/N_c)} 11.4702 \left[ 1 +  (c_1+2d_1)z^{-1} + O\left(z^{-2} \right) \right]  .
\end{equation}
The relationships among $c_n, d_n$ ($n \ge 2$) are given in Appendix where trick ending results 
$\gamma = 0 = \delta$ and $c_n =0=d_n$ are obtained from the consistency of the coupled SD equation. 

Another solution is obtained as follows.  
Both the gluon and ghost loops contribute equally to the leading order, only when $\alpha=0=\beta$.  
From the matching of the last term in the respective contribution, we obtain
\begin{align}
 \alpha = 0 = \beta, \quad \text{and} \quad
 -\gamma =2\delta+1 = 2\gamma+1, i.e., \gamma = -{1/3} = \delta .
\end{align}
This is the same result as the UV case of Atkinson and Bloch \cite{AB98}.
If $\alpha=0$, then $\beta=0$, and vice versa.
This solution implies the UV asymptotic freedom, although it leads to a different value $\beta_0={9 \over 4}N_c \tilde{Z}_1$ for the coefficient of the $\beta$ function from the perturbative result $\beta_0={11 \over 3}N_c \tilde{Z}_1$ where $\tilde{Z}_1=1$.  
This disagreement was resolved by taking more involved renormalization prescription in \cite{SHA97,FAR02}.  
The correction terms with coefficients $c_n, d_n$ can be obtained in a self-consistent way.  
The more details will be given in a subsequent paper.

\section{Conclusion and discussion}

We have proposed a new Ansatz for the asymptotic solution of the coupled SD equation for the gluon and ghost form factors in Yang-Mills theory.  
This Ansatz can be applied to find the UV and IR asymptotic solution simultaneously.  
In fact, we have reproduced the UV asymptotic solution with logarithmic behavior and IR asymptotic solution with power behavior.  
Moreover, we have pointed out a possibility that an IR asymptotic solution can have a  logarithmic correction to the power behavior found so far,
with keeping the IR fixed point.  
However, it is proved in Appendix that the logarithmic correction in the IR asympotic solution is excluded for a version of the truncated SD equation treated in this paper.  In this case, therefore, (\ref{IR}) is replaced by 
\begin{align}
  \alpha = - 2\beta \not= 0, \quad \gamma = 0 = \delta  , \quad
c_n = 0 = d_n . 
\label{IR2}
\end{align}
and only the power correction (\ref{ABsolution}) is allowed in the IR region, which was presumed in the previous papers \cite{SHA97,AB98}.  
This result is in sharp contrast with the UV asymptotic solution which allows power corrections to the logarithmic behavior \cite{Kondo02b}. 
 
 The issues to be clarified in subsequent works are as follows.

1)  We are to study the renormalization group properties associated with the IR asymptotic solution, e.g., the $\beta$ function and the anomalous dimension near the IR fixed point.  
They might be helpful to exclude the logarithmic correction in the IR asymptotic solution.  
The result will enable us to select the physical solution with renormalization group invariance, since all of the solutions of SD equation are not necessarily the physical solution. 

2) In this paper we have adopted the bare vertex function.  Therefore, it will be important to see whether our result remains true even after  the vertex function is improved so as to satisfy the Slavnov-Taylor identity.  

3) It will be interesting to study the IR asymptotic solution in other covariant gauges, e.g., the Maximal Abelian (MA) gauge \cite{tHooft81}, in addition to the non-covariant gauge,  e.g., Coulomb gauge \cite{SS01,CZ02}.   
We wish to know how the behavior of IR solution in the MA gauge is connected to the dual superconductor picture \cite{dualsuper} for quark confinement \cite{KondoI}.  

4) We have not yet worked out the effect of the dimensionality of space-time on IR solution.  For example, we can ask whether the presence or absence of the power or logarithmic correction is specific to the four-dimensional case or not.

5) It is not yet clear how the UV solution with gluon condensation \cite{GSZ01,Boucaudetal00,Kondo01} is connected to the IR solution in a quite narrow transition region.  Does the mass scale, e.g., the Gribov mass in IR region have something to do with the gluon condensation?

%%%%%%%%%%%%%%%%%%%%%%%%%%%%%%%%%%%%%%%%%%%%%%%%%%
%%%%%Acknowledgments
%%%%%%%%%%%%%%%%%%%%%%%%%%%%%%%%%%%%%%%%%%%%%%%%%%

\section*{Acknowledgments}
The author would like to thank Reinhardt Alkofer, Jacques C.R. Bloch, Kurt Langfeld and Hugo Reinhardt for kind hospitality in T\"ubingen University. 
He is grateful to Takahito Imai for drawing figures.  
This work is supported in part by Sumitomo Foundations and by 
Grant-in-Aid for Scientific Research from the Ministry of
Education, Science and Culture: (B)13135203 and (C)14540243.

\appendix 

\section{Proof of absence of logarithmic corrections in the IR limit}

Substituting the Ansatz (\ref{ansatz}) into the SD equation and performing the integration explicitly, we obtain the recursion relations by equating the coefficients of $z^{-N}$ in both sides:
\begin{subequations} 
\begin{align}
 & \left[ G_N - \sum_{n,m\ge 0:n+m=N} d_m c_n \right] 
e^{-{\alpha+2\beta \over \omega}z} z^{-(\gamma+2\delta)} 
\nonumber\\   
=& - {3N_c \over 4}\nu \sum_{\ell=1}^{N} \Big\{ 
{\omega^{\ell} \over (2+\beta)^{\ell+1}}  \sum_{n,m\ge 0:n+m=N-\ell}
 d_m c_n \prod_{i=0}^{\ell-1} (m-\delta+i)
\nonumber\\& \quad \quad \quad \quad \quad  
 - {\omega^{\ell} \over (\alpha+\beta)^{\ell+1}} \sum_{n,m\ge 0:n+m=N-\ell}
 d_m c_n \prod_{i=0}^{\ell-1} (m+n-\gamma-\delta+i) \Big\} ,
\label{Gint}
\\
 & \left[ F_N - \sum_{n,m\ge 0:n+m=N} d_m d_n \right] 
e^{-{\alpha+2\beta \over \omega}z} z^{-(\gamma+2\delta)} 
\nonumber\\  
=& {N_c \over 3}\nu \sum_{\ell=1}^{N} \Big\{ 
 \left[
 - {\omega^{\ell} \over (3+\beta)^{\ell+1}}  
+{3 \over 2} {\omega^{\ell} \over (2+\beta)^{\ell+1}}
\right]
\sum_{n,m\ge 0:n+m=N-\ell}
 d_m d_n \prod_{i=0}^{\ell-1} (m-\delta+i)
\nonumber\\& \quad \quad \quad \quad   
 - {1 \over 2}{\omega^{\ell} \over (2\beta)^{\ell+1}} \sum_{n,m\ge 0:n+m=N-\ell}
 d_m d_n \prod_{i=0}^{\ell-1} (m+n-2\delta+i) \Big\} ,
\label{Fint}
\end{align}
\end{subequations} 
where $G_N$ and $F_N$ are defined by
$
 F^{-1}(z) = A^{-1}e^{-\alpha z/\omega} z^{-\gamma} \sum_{N=0}^{M} F_N z^{-N} 
$
and
$
 G^{-1}(z) = B^{-1}e^{-\beta z/\omega} z^{-\delta} \sum_{N=0}^{M} G_N z^{-N} 
$
and explicit calculations show  
\begin{align}
 G_1 =& -d_1, \quad 
 G_2 = -d_2+d_1^2, \quad 
 G_3 = -d_3+2d_1 d_2-d_1^3 , \quad 
\nonumber\\   
 G_4 =& -d_4+2d_1 d_3-3d_1^2 d_2+d_1^4 , 
  \cdots , 
\nonumber\\
 F_1 =& -c_1,  \quad 
 F_2 = -c_2+c_1^2 , \quad 
 F_3 = -c_3+2c_1 c_2-c_1^3 , \quad 
\nonumber\\    
 F_4 =& -c_4+2c_1 c_3-3c_1^2 c_2+c_1^4 , 
 \cdots .
\end{align}
It is easy to find that
$\alpha+2\beta=0$ and $\gamma+2\delta=0$.

In the following, we shall prove $\gamma=0=\delta$ and $c_N=0=d_N$ for $N \ge 1$ by  mathematical induction. 
First, we show that $\gamma=0=\delta$. For $N=1$, the above recursion relations lead to 
\begin{subequations} 
\begin{align}
 -d_1 - (d_1+c_1) 
=& - {3N_c \over 4}\nu  \Big\{ 
{1 \over (2+\beta)^2 }   
 + {1 \over (\alpha+\beta)^2 }  
 \Big\}  \omega (-\delta) ,
\label{Gint1}
\\
  -c_1 - 2d_1 
=&  - {N_c \over 3}\nu  \Big\{ 
   {1 \over (3+\beta)^{2}}   
 -{3 \over 2} {1 \over (2+\beta)^{2}}    
 +  {1 \over (2\beta)^{2}}   
 \Big\}  \omega (-\delta) ,
\label{Fint1}
\end{align}
\end{subequations} 
where we have used $\gamma=-2\delta$ to derive (\ref{Gint1}). Eq.(\ref{Gint1}) and (\ref{Fint1})  agree with  
(\ref{ghost4}) and (\ref{gluon4}), respectively. 
Substituting $\alpha=-2\beta$ and $\beta=-0.769479$ into the RHS leads to  different coefficients of $N_c \nu \omega \delta$, although the LHS are the same, i.e., $c_1+2d_1=-1.762N_c \nu \omega \delta$ and $0.12247N_c \nu \omega \delta$. 
Therefore, we conclude $\delta=0$ and hence $\gamma=0$.  Thus we obtain $c_1+2d_1=0$ or $c_1=-2d_1$. 
\par
For $N=2$, we use $\gamma=0=\delta$ to write the recursion relations, 
\begin{subequations} 
\begin{align}
  -d_2+d_1^2 - (d_2+d_1c_1+c_2) 
=& - {3N_c \over 4}\nu  \Big\{ 
{1 \over (2+\beta)^{2}}    
 + {1 \over (\alpha+\beta)^{2}}   
   \Big\} \omega d_1  ,
\\
  -c_2+c_1^2 - (d_2+d_1^2+d_2) 
=& - {N_c \over 3}\nu  \Big\{ 
   {1 \over (3+\beta)^{2}}   
 -{3 \over 2} {1 \over (2+\beta)^{2}}  
 +  {1 \over (2\beta)^{2}}   
\Big\} \omega d_1 .
\end{align}
\end{subequations} 
 For $c_1=-2d_1$, both LHS agree with  
$-c_2-2d_2+3d_1^2$.  However, the coefficients of $N_c \nu \omega d_1$ in RHS are not equal to each other.  Hence, we obtain $d_1=0$ and $c_1=0$. 
Thus, we obtain $c_2+2d_2=0$. 
\par
 For $N \ge 3$, suppose that $c_m=0=d_m$ for $m=1,2, \cdots, N-2$ and 
$c_{N-1}+2d_{N-1}=0$ (Note that $c_0=1=d_0$).
Then, the non-vanishing contributions in RHS of (\ref{Gint}) or (\ref{Fint}) stem from the terms containing the non-zero $c_{N-1}$ or $d_{N-1}$.  Hence it is sufficient to consider the $\ell=1$ part in RHS.  
Thus, the recursion relations reduce to
\begin{subequations} 
\begin{align}
  G_N - (d_N+c_N) 
=& - {3N_c \over 4}\nu \Big\{ 
{\omega \over (2+\beta)^{2}} 
d_{N-1} 
 - {\omega \over (\alpha+\beta)^{2}}  
  (d_{N-1}+c_{N-1}) 
\Big\} (N-1)
\nonumber\\
=& - {3N_c \over 4}\nu \Big\{ 
{1 \over (2+\beta)^{2}} 
 + {1 \over (\alpha+\beta)^{2}}  
\Big\} (N-1)d_{N-1} \omega 
,
\label{Gint2}
\\
  F_N -  2d_N  
=&
 -{N_c \over 3}\nu \Big\{ 
 {1  \over (3+\beta)^{2}}   
 -{3 \over 2} {1  \over (2+\beta)^{2}} 
 +  {1 \over (2\beta)^{2}} 
  \Big\}   (N-1)d_{N-1} \omega .
\label{Fint2}
\end{align}
\end{subequations} 
Under the assumption,
we have the simple expression, 
$G_N=-d_N$ and $F_N=-c_N$. 
The LHS of both equations reduce to $-(c_N+2d_N)$.  
By substituting the value of $\beta=-\kappa$ into the RHS, we find that the respective RHS gives the different coefficient for 
$N_c \nu (N-1)d_{N-1}\omega$. This implies that $d_{N-1}=0$ and hence $c_{N-1}=0$. 
Thus we obtain
$c_N+2d_N=0$.
Repeating this argument, we obtain 
$c_N=0=d_N$ and $c_{N+1}+2d_{N+1}=0$. 
Thus we can conclude 
$c_N=0=d_N$ for all $N \ge 1$.

%\newpage
%\baselineskip 12pt

\end{document}